\documentclass[journal]{IEEEtran}
\usepackage{graphicx,amssymb,lineno}
\usepackage{amsmath,amsfonts,amssymb}

\usepackage{algorithmic}
\usepackage[usenames]{color}
\usepackage{float}
\usepackage{cite}
\usepackage{stfloats}

\usepackage[linesnumbered,ruled,vlined]{algorithm2e}

\usepackage{graphicx,graphics,color,epsfig,subfigure,graphpap,rotate}
\usepackage{times, verbatim, subfigure, epsfig, graphicx, latexsym, amsmath}
\usepackage{url}
\usepackage{subfigure}

\begin{document}

\title{Device-to-Device Communications Enabled Multicast Scheduling with
 the Multi-Level Codebook in mmWave Small Cells}

\author{Yong~Niu,
        Liren~Yu,
        Yong~Li,~\IEEEmembership{Senior Member,~IEEE},
        Zhangdui Zhong,
        Bo Ai,
        and Sheng~Chen,~\IEEEmembership{Fellow,~IEEE}
\thanks{Y. Niu, Z. Zhong, and B. Ai are with State Key Laboratory of Rail Traffic Control
 and Safety, Beijing Engineering Research Center of High-speed Railway Broadband
 Mobile Communications, School of Electronic and Information Engineering, Beijing
 Jiaotong University, Beijing 100044, China (E-mails: niuy11@163.com; zhdzhong@bjtu.edu.cn).} %
\thanks{L.~Yu and Y. Li are with Tsinghua National Laboratory for Information Science and Technology
 (TNLIST), Department of Electronic Engineering, Tsinghua University, Beijing 100084, China (E-mails:
 ylr14@mails.tsinghua.edu.cn, liyong07@tsinghua.edu.cn).} %
\thanks{S. Chen is with School of Electronics and Computer Science, University of
 Southampton, Southampton SO17 1BJ, UK (E-mail: sqc@ecs.soton.ac.uk), and
 also with King Abdulaziz University, Jeddah 21589, Saudi Arabia.} %
\vspace*{-5mm}
}

\maketitle

\begin{abstract}
 With the exponential growth of mobile data, there are increasing
 interests to deploy small cells in millimeter wave (mmWave) bands to underlay the
 conventional homogeneous macrocell network as well as in exploiting device-to-device
 (D2D) communications to improve the efficiency of the multicast service that
 supports content-based mobile applications. To compensate for high propagation loss
 in the mmWave band, high-gain directional antennas have to be employed, while it is
 critical to optimize multicast service in order to improve the network performance.
 In this paper, an efficient multicast
 scheduling scheme is proposed for small cells in the mmWave band, called MD2D, where both D2D
 communications in close proximity and multi-level antenna codebook are utilized.
 Specifically, a user partition and multicast path planning algorithm is proposed to
 partition the users in the multicast group into subsets and to determine the
 transmission node for each subset, so as to achieve optimal utilization of D2D
 communications and multi-level antenna codebook. Then a multicast scheduling
 algorithm schedules the transmission for each subset. Furthermore, in order to
 optimize the network performance, the optimal choice of user partition thresholds is
 analyzed. Performance evaluation demonstrate that the
 MD2D achieves the best performance, in terms of network throughput and energy
 efficiency, compared with other existing schemes. MD2D improves
 the network throughput compared with the second-best scheme by about 27\%.
\end{abstract}


\section{Introduction}\label{S1}

 Mobile traffic demands are explosively increasing over the past decade \cite{data2}.
 In order to improve mobile network capacity accordingly, small cells underlying the
 macrocell have been proposed and received much
 attention, which is referred to as heterogeneous
 cellular networks (HCNs) \cite{Pico_60GHz}. By utilizing the millimeter wave (mmWave) bands, and by a dense deployment of small
 cells \cite{Pico_60GHz,dense-cells}, HCN can significantly improve the network
 capacity with less interference, compared to a conventional HCN deployment.
 The mmWave bands, such as the 28\,GHz band, the 60\,GHz band, and the E-band, have abundant spectrum. Consequently, multi-Gbps communication services can be supported \cite{elkashlan2015millimeter}. Moreover,
 rapid development in complementary metal-oxide-semiconductor technology for radio frequency
 integrated circuits paves the way for electronic products in the mmWave band
 \cite{rappaport2011state}. Several standards have been defined for indoor wireless
 personal area networks (WPAN) and wireless local area networks (WLAN), such as
 IEEE 802.15.3c \cite{IEEE_802.15.3c}, IEEE 802.11ad \cite{IEEE_802.11ad}, and IEEE 802.11ay \cite{802.11ay}.


 With higher carrier frequency, mmWave
 communications suffer from higher propagation loss. Thus, directional antennas
 need to be adopted at both transmitter and receiver to form directional high-gain
 beams in order to compensate the high propagation loss \cite{beam_training,Xiao1,Xiao2,Xiao3}.
 It is well-known that wide beams have low antenna gains and can only support low
 transmission rates, while narrow beams have high antenna gains and can support high
 transmission rates. With the aid of the multi-level antenna codebook, therefore,
 transmitters can use wide beams to communicate with multiple low-rate receivers, and
 use narrow beams to communicate at a high transmission rate. Hence, by optimizing the
 beam selection in a multi-level antenna codebook, the flexibility of directional
 beams can be utilized to enhance the achievable network throughput.

 It is found that a small amount of popular content occupy the
 majority of requests in content downloading, and the content popularity in mobile networks is found
 to obey the Zipf's law \cite{content_popularity}. In multicast service, the access
 point (AP)
 provides multiple users within a multicast group with the same data
 \cite{pcds,sdm}. To serve more users simultaneously, wider
 beams are preferred. But wider beams with lower antenna gain can only support lower
 transmission rate, which degrades multicast efficiency. Thus,
 dividing the multicast group into multiple subsets becomes necessary. On the other hand, in the user-intensive region,
 where the multicast service is usually applied, two
 user devices will be probably located near to each other. In this case, device-to-device (D2D)
 communications can be enabled for improving multicast
 efficiency, owing to better channel conditions \cite{Yong}. For the multicast traffic, users already with the multicast content can use the D2D communications to forward the content to users nearby. Thus, D2D communications usually have shorter distance, and thus the propagation loss is less. Consequently, higher transmission rate can be obtained, and thus less transmission time is needed. Thus, network throughput can be increased. With less time needed to accommodate the multicast demands, lower energy consumption can be achieved if the transmission power is fixed. Therefore, the energy efficiency can be increased.

 Against the above background, in this paper, we investigate the problem of optimal multicast
 scheduling in mmWave small cells underlaid by D2D communications. This problem is
 challenging because to serve users efficiently, users in the multicast group must be
 partitioned optimally into subsets and beams must be selected optimally to serve users
 in each subset. Moreover, user device with the multicast traffic must be capable of
 serving other subsets efficiently by exploiting better D2D channels. The D2D communication here is for multicast transmission, and ``D2D'' here means ``one device to multiple devices using directional beams'', which is very different from previous works. To obtain a
 practical solution, we develop an efficient multicast scheduling scheme, called
 MD2D, where appropriate beams are selected to serve users efficiently in each multicast
 transmission, while D2D communications are utilized to improve multicast efficiency.
 The contribution is three-fold as summarized below.
\begin{itemize}
\item We formulate the problem of the optimal multicast scheduling with D2D communications
 and beam selection in a multi-level codebook considered into a mixed integer nonlinear program that
 minimizes the total multicast transmission time, by efficient multicast group partition,
 beam selection, and D2D communication utilization.
\item To obtain a practical solution to this challenging problem, an efficient
 multicast scheduling scheme is proposed, called MD2D, where two algorithms are proposed, user partition
 and multicast path planning, and multicast scheduling. The first algorithm appropriately
 partitions the users in the multicast group into subsets and determines the transmission
 node for each subset, while the second one schedules the transmission for each subset
 efficiently.
\item We further investigate the optimal selection of user partition thresholds to
 optimize the achievable network performance. Extensive evaluations under various system
 settings show our proposed MD2D achieves the best performance, in
 network throughput and energy efficiency, compared with other
 schemes.
\end{itemize}

 The structure of this paper is as follows. Section~\ref{S2} reviews the related
 work on the media access control (MAC) protocols for mmWave small cells.
 Section~\ref{S3} illustrates the system model and the basic idea of our
 MD2D. Section~\ref{S4} formulates the optimal multicast scheduling
 problem with D2D communications, multicast group partition and beam selection in a
 multi-level codebook. Section~\ref{S5} is entirely devoted to our proposed multicast scheduling
 scheme, namely, MD2D. Section~\ref{S6} evaluates the performance of our MD2D scheme,
 in terms of network throughput and energy consumption, using there existing schemes as
 the benchmarks.  Section~\ref{S7} gives the conclusion.

\section{Related Work}\label{S2}

 There exist some related works on directional MAC protocols for WPANs and WLANs in the
 mmWave band \cite{Qiao,EX_Region,Qiao_6,Qiao_15,Qiao_7}. In WPANs and WLANs, time
 division multiple access (TDMA) protocol is traditionally adopted \cite{ECMA_387,IEEE_802.15.3c}.
 Cai \emph{et al.} \cite{EX_Region} derived the conditions of exclusive region that
 concurrent transmission always outperforms TDMA. In an indoor IEEE 802.15.3c WPAN, the work of \cite{Qiao} proposed a concurrent
 transmission scheduling algorithm to maximize the number of flows with the quality
 of service requirement for each flow satisfied. In \cite{Qiao_7}, a
 multi-hop concurrent transmission scheme is developed to overcome the link outage problem and to enhance flow throughput.
 For TDMA based protocols, an unfair
 medium time allocation problem exists for individual users under bursty data traffic
 \cite{mao}.

 There also exist some centralized protocols proposed for WPANs or WLANs in the mmWave band
 \cite{Gong,MRDMAC,mao,chenqian}. In \cite{MRDMAC}, a multihop-relay based
 directional MAC (MRDMAC) protocol is proposed, where the PNC applies a weighted round robin scheduling
 to overcome the deafness problem. MRDMAC utilized multi-hop relaying to overcome blockage. Son \emph{et al.}
 \cite{mao} proposed a frame based directional MAC protocol (FDMAC). The core of this FDMAC is the greedy coloring algorithm, which utilizes concurrent transmissions
 to improve the network throughput. Niu \emph{et al.} \cite{tvt_own} proposed a blockage-robust
 and efficient directional MAC (BRDMAC) protocol to overcome the blockage problem by two-hop
 relaying. Recently, Niu
 \emph{et al.} \cite{JSAC_own} proposed a joint transmission scheduling protocol for the radio
 access and backhaul of small cells in mmWave band, called D2DMAC. In D2DMAC, a path selection
 criterion is proposed to exploit D2D transmissions when performance improvement is available.
 Zhang \emph{et al.} \cite{Haijun} investigate user association and power allocation in mmWave-based
 ultra dense networks with attention to load balance constraints, energy harvesting by base stations,
 user quality of service requirements, energy efficiency, and cross-tier interference limits. To solve
 the joint user association and power optimization problem, they proposed an iterative gradient user association
 and power allocation algorithm to achieve an optimal point.

 In terms of multicast communication, there also exist a few works on MAC protocols for WPANs
 and WLANs in the mmWave band \cite{sdm,IMG}. Naribole \emph{et al.} \cite{sdm} implemented a technique called
 scalable directional multicast (SDM) to train the AP with per-beam per-client
 received-signal-strength-indicator measurements via partially traversing a codebook tree.
 Based on the training information, they proposed a scalable beam grouping algorithm to obtain
 the minimum multicast group data transmission time. Park
 \emph{et al.} \cite{IMG} proposed an incremental multicast grouping (IMG) scheme where the
 beamwidths are adaptively assigned. However, D2D communications were not enabled in this IMG.


 It is clear that jointly utilizing multi-level codebook and D2D communications to maximize
 multicast efficiency for small cells in the mmWave band is challenging and has not been
 exploited in the open literature.

\section{System Overview}\label{S3}

\subsection{System Model}\label{S3-1}

 In a mmWave small cell of $n$ nodes, one node is the AP and the rest
 are user equipments (UEs). The system time is partitioned into non-overlapping time slots of
 equal length. The AP synchronizes the clocks of UEs and schedules the medium
 access to accommodate the multicast demand. Equipped with steerable
 directional antennas, the AP and users generate the directional beams of different beamwidths
 via a multi-level codebook. We assume that a bootstrapping program is run in the system so
 that the AP has the network topology and the location information of UEs
 \cite{bootstrapping,location_1,mmW-V2V}.

We assume the mmWave small cells are deployed
underlying the macrocell to form the heterogeneous cellular network (HCN), and the small-cell APs and UEs are also equipped
with omnidirectional antennas for 4G communications. Thus, the location information
can also be obtained by the localization techniques in the cellular bands.
Meanwhile, the transmission requests and some signaling information
for mmWave small cells can be collected by the reliable 4G communication.


 Because non-line-of-sight
 transmissions suffer from very high attenuation, mmWave communications in small cell mainly
 rely on line-of-sight (LOS) transmissions. Therefore, we assume that a LOS path is available
 for each transmission \cite{NLOS}.

 Denote the directional link from node $i$ to $j$ by $(i,j)$. In directional beamforming,
 both nodes $i$ and $j$ point toward each other via a beam from an $L$-level codebook.
 Assume that transmit node $i$ adopts the $t$th beam in the $l$th level of the codebook, which
 is denoted as $\varphi (t,l)$, and the antenna gain of $\varphi (t,l)$ in the direction of
 $i\to j$ is $G_{ij}^{(\rm{T})}(\varphi (t,l))$, while receive node $j$ adopts the $s$th beam
 in the $h$th level of the codebook, which is denoted by $\varphi (s,h)$, and the antenna gain
 of $\varphi (s,h)$ in the direction of $i\to j$ is $G_{ij}^{(\rm{R})}(\varphi (s,h))$. Then
 based on the path loss model \cite{NLOS}, the received power at node $j$ for link $(i,j)$
 is given by
\begin{equation}\label{eq1}
 P^{(\rm{R})}_{ij} = k_0 G_{ij}^{(\rm{T})}(\varphi (t,l)) G_{ij}^{(\rm{R})}(\varphi (s,h))
  d_{ij}^{-\tau} P_t ,
\end{equation}
 where $P_t$ is the transmission power and $k_0$ is a constant that is proportional to
 $\big(\frac{\lambda}{4\pi}\big)^2$ ($\lambda$ is carrier wavelength), while
 $d_{ij}$ is the distance between transmitter $i$ and receiver $j$ and $\tau$ is the path
 loss exponent \cite{Qiao}. Hence the received signal to noise ratio (SNR) of link $(i,j)$
 is calculated according to
\begin{equation}\label{eq2}
 \hspace*{-2mm}{\rm SNR}_{ij} = \frac{P^{(\rm{R})}_{ij}}{N_0 W}
  = \frac{k_0 G_{ij}^{(\rm{T})}(\varphi (t,l)) G_{ij}^{(\rm{R})}(\varphi (s,h))
  d_{ij}^{-\tau} P_t}{N_0 W} , \!
\end{equation}
 where $W$ is the bandwidth and $N_0$ is the one-sided power spectra density of
 the link's white Gaussian noise \cite{Qiao}. Considering the reduction of multipath effect for
 directional mmWave links \cite{MRDMAC}, the achievable data rate of link $(i,j)$ can be
 estimated based on Shannon's channel capacity as
\begin{align}\label{model_rate} 
 & \hspace*{-2mm}R_{ij} = \eta W \log_2\big(1 + {\rm SNR}_{ij}\big) ,
\end{align}
 where $\eta\in (0,1)$ denotes the efficiency of the transceiver \cite{Qiao}.

\begin{figure}[tp]
\vspace*{-1mm}
\begin{minipage}[t]{1\linewidth}
\centering
\includegraphics[width=0.95\columnwidth]{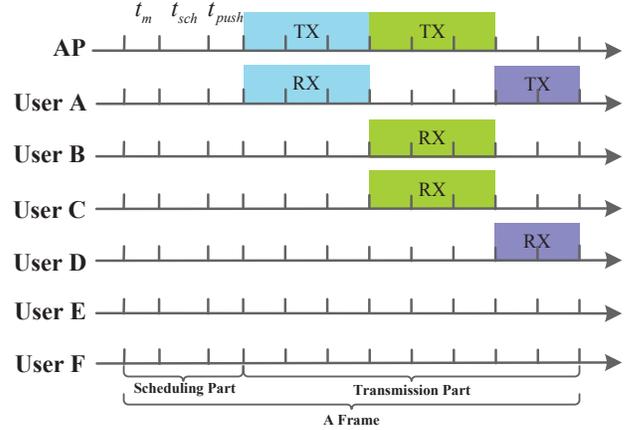}
\centerline{\small (a) Time-line illustration of MD2D}
\end{minipage}%
\vfill\vspace{0.6cm}
\begin{minipage}[t]{1\linewidth}
\centering
\includegraphics[width=0.85\columnwidth]{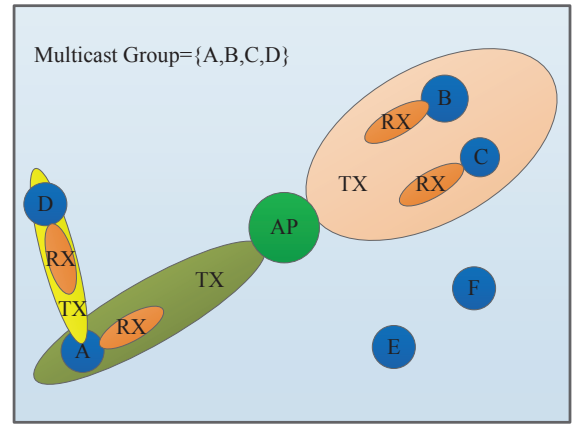}
\centerline{\small (b) Network topology and MD2D operation}
\end{minipage}
\caption{An example of MD2D operation in a small cell, with the multicast group
 consisting of users A to D.}
\label{fig:MD2D operation} 
\vspace*{-3mm}
\end{figure}

\subsection{Problem Overview}\label{S3-2}

 We consider the multicast service transmitted from the AP to a multicast group. To
 improve multicast efficiency, appropriate beams should be selected from the multi-level
 codebook for each multicast transmission. How to select appropriate beams is illustrated in Section \ref{S5-1}.
 Besides, users should be able to receive the multicast service from nearby users through D2D communications.
 To illustrate our MD2D scheme, which exploits both D2D transmissions and multi-level
 codebook, we use the simple example depicted in Fig.~\ref{fig:MD2D operation}, which is
 a small cell of six users, with the multicast group consisting of users A to D.

 Fig.~\ref{fig:MD2D operation}\,(a) depicts the time-line illustration of this example. In
 the system, time is divided into a sequence of non-overlapping frames \cite{mao}. Each
 frame has two periods, multicast scheduling period and multicast transmission period.
 In the scheduling period, the AP first obtains the multicast traffic and the information of
 the associated multicast group from the network layer, which needs time $t_m$; next the AP
 computes a schedule required to complete the multicast service, which needs time $t_{sch}$;
 then AP pushes the schedule to the users in sequence, which needs
 time $t_{push}$. In the transmission period, all the nodes in the multicast group begin
 transmissions according to the schedule until the multicast traffic is distributed to all the
 users in the group. As illustrated in Fig.~\ref{fig:MD2D operation}\,(a), a multicast
 transmission period is naturally divided into multiple phases according to the schedule,
 and in each phase, the multicast transmission occupies several time slots.

 Fig.~\ref{fig:MD2D operation}\,(b) illustrates the network topology of this simple example.
 To serve this multicast group efficiently, the multicast group can be partitioned into
 three subsets, namely, user A, users B and C, and user D. Since users B and C are
 located sufficiently close with a very small angle difference, the AP can serve them with a
 wide beam simultaneously. By contrast, since user D and user A span a large angle, serving
 them simultaneously with a wide beam will lead to low transmission rate and, consequently,
 requires many more time slots. Instead, the AP may serve user A with a narrow beam, and
 then by exploiting the close proximity of users A and D, the AP can let user A to serve user
 D via the D2D communication.

 Thus, the schedule as shown in Fig.~\ref{fig:MD2D operation}\,(a) is obtained, which consists
 of three multicast transmission phases in a frame. In the first phase, the AP directs its
 narrowest beam towards user A to achieve the highest transmission rate, as illustrated in
 Fig.~\ref{fig:MD2D operation}\,(b), which occupies three time slots. In the second phase,
 AP serves users B and C simultaneously with a wide beam, which occupies three time slots.
 Finally, user A serves user D with the narrowest beam to achieve highest transmission rate
 in the third phase. Because users A and D are very close, this D2D based third phase only
 occupies two time slots. Thus, a total of eight time slots are needed to serve the multicast
 group. By contrast, if the AP serves the four users sequentially with the narrowest beams, a
 total of twelve time slots are needed.

 From this simple example, we can clearly observe that there are two key problems to solve in
 order to maximize multicast efficiency. The first one is how to partition the multicast group
 into subsets and serve each subset via an appropriate beam to achieve high multicast efficiency.
 The second one is how to effectively utilize D2D communications by exploiting the physical
 proximity of devices to further improve multicast efficiency as much as possible.

\section{Problem Formulation}\label{S4}

 The set of users in the multicast group is denoted by $\mathbb{U}$ and partition $\mathbb{U}$
 into $S$ subsets, i.e., $\mathbb{U}=\mathbb{U}_1\cup\mathbb{U}_2\cdots \mathbb{U}_S$. Note
 that the users in each subset receive multicast service simultaneously. We denote the $j$th
 user in the $i$th subset $\mathbb{U}_i$ by $u_{ij}$. Since $|\mathbb{U}_{i}|$ is at least 1,
 $1 \le S \le |\mathbb{U}|$. The traffic demand for the multicast group is denoted by $D$. The
 schedule for the multicast transmission period of a frame contains $K$ phases, and each phase
 lasts several consecutive time slots.

 For each phase, we define the size $1\times S$ vector ${\bf a}^k$ to indicate whether the
 subsets of the multicast group are scheduled in the $k$th phase, where $1\le k\le K$.
 Specifically, let the $i$th element of ${\bf a}^k$ be denoted by $a_i^k$, where $1\le i\le S$.
 If $\mathbb{U}_i$ is scheduled in the $k$th phase, $a_i^k=1$; otherwise, $a_i^k=0$. We denote
 the transmit node for the subset $\mathbb{U}_i$ by $s_i$. Since D2D communication is enabled,
 $s_i$ may be the AP or a user with the multicast data. When $s_i$ is a user with the multicast
 data, we denote the multicast subset that $s_i$ belongs to by $\mathbb{U}_{f(s_i)}$, i.e.,
 $f(s_{i})$ denotes the subset number of $s_i$. For the user $u_{ij}\in \mathbb{U}_i$, the
 achievable transmission rate provided by $s_{i}$ is $R_{ij}$ as given in (\ref{model_rate}),
 which is rewritten here
\begin{align}\label{eq4}
 & \hspace*{-1mm}R_{ij} \! =\! \eta W \log_2\! \Big(\! 1\! + \! \frac{k_0 G_{s_iu_{ij}}^{(\rm{T})}(\varphi (t,l))
  G_{s_iu_{ij}}^{(\rm{R})}(\varphi (s,h)) d_{s_iu_{ij}}^{-\tau} P_t}{N_0 W}\Big) \! ,
\end{align}
 where again $\varphi (t,l)$ denotes the transmit beam of $s_i$ and $\varphi (s,h)$ is the receive
 beam of $u_{ij}$. We denote the set of beams, i.e., the codebook with $L$ levels, by $\mathbb{C}_L$.
 Therefore, $\varphi (t,l),\varphi (s,h)\in \mathbb{C}_L$. We further denote the required
 transmission rate to serve the users in $\mathbb{U}_{i}$ simultaneously by $R_i$. We denote
 the number of time slots scheduled for the $k$th phase by $\delta^k$, and the time slot duration is $\Delta$.

 To maximize the multicast efficiency or throughput, the transmission schedule should
 accommodate the multicast demand of all the users with a minimum number of time slots. Therefore,
 the objective function to be minimized is simply
\begin{align}\label{eq5}
  \mathcal{J} =& \sum\limits_{k=1}^K \delta^k .
\end{align}
 Next we analyze the system constraints of this multicast transmission optimization
 problem.

 First, each multicast group is partitioned into several subsets, which is expressed as
 the following two constraints
\begin{align} 
 \mathbb{U} =& \mathbb{U}_1\cup\mathbb{U}_2\cup\cdots \cup\mathbb{U}_S , \label{cons1}  \\
 & 1 \le S \le |\mathbb{U}| . \label{cons2}
\end{align}
 Second, since $a_{i}^k$ is a binary variable, we have the constraint
\begin{equation}\label{cons3}  
 a_i^k =\{0,1\} , ~ \forall i\in\{1,2,\cdots ,S\} , \forall k\in\{1,2,\cdots ,K\} .
\end{equation}
 Third, to reduce beamforming overhead and system complexity, we restrict to the case that the multicast
 transmission for each subset is only scheduled once in one frame of the schedule.
 Thus, we have
\begin{equation}\label{cons4} 
 \sum\limits_{k=1}^K a_i^k = 1 , \ \ \forall i\in\{1,2,\cdots ,S\} .
\end{equation}
 Fourth, to serve the users in each subset simultaneously, the required transmission
 rate must meet the condition
\begin{equation}\label{cons5} 
 R_i = \min\limits_{u_{ij}\in \mathbb{U}_i} R_{ij}, \ \ \forall i\in\{1,2,\cdots ,S\} ,
\end{equation}
 where $R_{ij}$ is given by (\ref{eq4}) with the constraint
\begin{equation}\label{cons52} 
 \forall \varphi (t,l)\in \mathbb{C}_L, ~ \forall \varphi (s,h)\in \mathbb{C}_L .
\end{equation}
 Fifth, the schedule must meet the multicast demand, and therefore we must have
\begin{equation}\label{cons6}  
 \sum\limits_{k=1}^K \big( a_i^k \cdot R_i \cdot \delta^k \cdot \Delta\big) \ge D ,
  \ \ \forall i\in\{1,2,\cdots ,S\}.
\end{equation}
 Lastly, to be able to exploit D2D communications, the transmit node of each subset
 should obtain the multicast data first. Thus, if a subset has a user that transmits
 multicast data to another subset, then the multicast transmission to this subset
 should be scheduled prior to the D2D based  multicast transmission to the other subset.
 This constraint can be expressed as
\begin{equation}\label{cons7}  
 \sum\limits_{k=1}^{K^*} a_{f(s_i)}^k \ge \sum\limits_{k=1}^{K^*} a_i^k , \ \
  \forall i , \ K^*\in\{1,\cdots, K\} .
\end{equation}

 Thus, the optimal multicast scheduling problem (P1) can be expressed as follows
\begin{equation}\label{obj} 
\begin{array}{ll}
 ({\rm P1}) & \min \mathcal{J} = \sum\limits_{k=1}^K \delta^k , \\
 & \text{s.t. Constraints } (\ref{cons1}) \text{ to } (\ref{cons7}) \text{ are met.}
\end{array}
\end{equation}
 Note that constraints (\ref{cons5}) and (\ref{cons6}) are nonlinear, and there
 exists set partitioning operation in constraint (\ref{cons1}). Thus the problem
 is a mixed integer nonlinear programming, where $a_i^k$ are binary variables,
 $\delta^k$ are integer variables and $S$ is an integer variable, while $\varphi (t,l)$
 and $\varphi (s,h)$ are discrete variables. This problem is more complex than the
 NP-complete 0-1 Knapsack problem \cite{Qiao,Knapsack}. The optimal solution can be obtained via the exhaustive search, which has high computational complexity, and cannot be applied in practice.

\section{Proposed Multicast Scheduling Scheme}\label{S5}

 As stated previously, there are two key mechanisms to improve multicast efficiency. First,
 the potential of multi-level codebook should be unleashed. Wide beams are able to cover
 larger angle range and may serve more users simultaneously. However, wide beams have low
 antenna gains, and achievable transmission rates may be low. By contrast, narrow beams
 have high antenna gains and are able to support high transmission rates. But narrow beams
 have limited coverage in angle range and may not be able to serve many users simultaneously.
 Second, the advantages of D2D communications in physical proximity should be reaped.
 If the users are located sufficiently close, multicast transmission using a wide beam may
 be more efficient. Clearly, optimizing the network performance based on these two
 mechanisms is a complex problem, which requires elaborate design for user partition,
 multicast path planning and beam selection for multicast transmission.
 To reduce complexity and achieve practical solutions, we propose the heuristic multicast
 scheduling scheme, MD2D, for the optimization problem (P1). Specifically, we first propose
 a user-clustering and multicast-path-planning algorithm to partition the multicast group
 into appropriate subsets and to decide the multicast transmission paths for the multicast
 group, which is required by constraints (\ref{cons1}) and (\ref{cons2}).


\subsection{User Partition and Multicast Path Planning}\label{S5-1}

 We start from the AP to find the nearest user subset. Users that are located very close to
 each other and span a limited angle range will be put into a subset to be served
 simultaneously. In this way, we can realize the potential of multi-level codebook, and
 use wide beams to serve more users. We continue to expand the set of the already allocated
 subsets including the AP by finding the possible user subset nearby. If such a newly selected
 subset is very close to one of the allocated subsets, then we enable the D2D communications
 between these two subsets to serve this new subset more efficiently. Of course, the multicast
 transmission to this new subset should be scheduled behind the transmission to the allocated
 subset. In this way, the advantages of D2D communications are exploited to enhance multicast
 efficiency. In other words, the objective function in problem (P1) can be optimized via D2D communications.

 Let us denote the AP in the small cell as $\mathbb{U}_0$ and the subset of users that the
 algorithm allocates in the $t$th iteration by $\mathbb{U}_t$. We also denote the set of the
 subsets that have been allocated and thus are able to serve other unallocated users by
 $\mathbb{U}_{\mathcal{M}}$. Since the AP has the multicast data to serve users, it can be
 regarded as a subset that has been allocated, and thus $\mathbb{U}_{\cal M}$ is initialized
 to $\mathbb{U}_{\cal M}=\{\mathbb{U}_0\}$. The allocated multicast transmission path from
 allocated subset $\mathbb{U}_s$ to the new subset $\mathbb{U}_t$ in the $t$th iteration is denoted
 by $\mathbb{P}^t$. For each subset $\mathbb{U}_t\in
  \mathbb{U}_{\cal M}$ with $t > 0$, we define the polar coordinates of user $i$ relative to
 the center of $\mathbb{U}_t$ by $\big(r_i^t,\theta_i^t\big)$. For each subset $\mathbb{U}_t$,
 we denote its transmit node by $s_t$, and the beam selected for $s_t$ to serve $\mathbb{U}_t$
 is denoted by $\varphi (b_t,l_t)$. We assume that all the users in $\mathbb{U}_t$ point to
 $s_t$ with the narrowest or finest receive beams. The radius threshold and angle threshold
 for user partition are denoted by $r_{th}$ and $\theta_{th}$, respectively.

\begin{algorithm}[tp!]
\caption{User Partition and Multicast Path Planning.}
\label{alg: multicast-D2D} 
 \textbf{Input}: the multicast group $\mathbb{U}$; \\
 \textbf{Initialization}:  $t$=0; $\mathbb{U}_{\cal M}=\{\mathbb{U}_0\}$; \\
 \While {$|\mathbb{U}| > 0$}
  {
   $t=t+1$;
   $\mathbb{U}_t=\emptyset$; $\mathbb{P}^t=\emptyset$; \\
   \For {{\rm{each}} $\mathbb{U}_s\in \mathbb{U}_{\cal M}$}
    {
     Find the user $i$ in $\mathbb{U}$ with the minimum $r_i^s$; \\
     $r^s=r_i^s$;
    }
   Find the subset $\mathbb{U}_s\in \mathbb{U}_{\cal M}$ with the minimum $r^s$; \\
   \For {{\rm{each user}} $j\in \mathbb{U}$}
    {
     Find the maximum angle difference $\theta_{\max}=\max\limits_{i\in \mathbb{U}_t} |\theta_j^s - \theta_i^s|$; \\
     \If {$\mathbb{U}_t=\emptyset$}
      {
       $\theta_{\max}=0$; \\
      }
     \If {$|r_j^s-r^s| \le r_{th}$ {\rm{and}} $\theta_{\max}\le \theta_{th}$}
      {
       $\mathbb{U}_t=\mathbb{U}_t \cup j$; \\
       $\mathbb{U}=\mathbb{U} - j$; \\
      }
    }
   $\mathbb{U}_{\cal M}=\mathbb{U}_{\cal M} \cup \{\mathbb{U}_t\}$;
   $\mathbb{P}^t=\{\mathbb{U}_s\to \mathbb{U}_t\}$; \\
   \For {{\rm{each node}} $i\in \mathbb{U}_s$}
    {
     Obtain $R_t^i$, its maximum achievable rate to serve $\mathbb{U}_t$; \\
     Obtain $\varphi (b_t^i,l_t^i)$, its corresponding selected beam to serve $\mathbb{U}_t$; \\
    }
    $s_t=\arg \max\limits_{i \in \mathbb{U}_s} R_t^i$; \\
    $\varphi (b_t,l_t)=\varphi (b_t^{s_t},l_t^{s_t})$; \\
   }
 \textbf{Return} $\mathbb{U}_{\cal M}$, $\mathbb{P}^t$, $s_t$ and $\varphi (b_t,l_t)$ for each $\mathbb{U}_t$.
\end{algorithm}

 The pseudo-code of this user partition and multicast path planning is listed in
 Algorithm~\ref{alg: multicast-D2D}. Starting from line~3, it iteratively partitions
 the users in $\mathbb{U}$ into subsets and schedules the multicast transmission for
 each subset until all the users are scheduled. Specifically, in the $t$th iteration,
 we first find a user with the shortest distance from a subset $\mathbb{U}_s\in
 \mathbb{U}_{\cal M}$ in lines 5--8. Then we allocate the users that are close to this
 user into the subset $\mathbb{U}_t$ in lines 9--15. We measure the closeness in
 terms of distance and angle with respect to the reference subset $\mathbb{U}_s$
 identified in lines 5--8. The angle that the current users span is denoted by
 $\theta_{\max}$, as indicated in lines 10--12. As shown in line 13, the users
 selected should be located not far from the reference radius $r^s$ by a threshold
 $r_{th}$, and the angle that the current users in the subset span after the candidate
 user $j$ is added should be no more than a threshold $\theta_{th}$. If the
 candidate user $j$ meets these two conditions, it is added into $\mathbb{U}_t$ and
 also removed from $\mathbb{U}$, as shown in lines 14--15. In line 16, the newly
 allocated subset $\mathbb{U}_t$ is added to $\mathbb{U}_{\cal M}$ and the
 multicast transmission from $\mathbb{U}_s$ to $\mathbb{U}_t$ is recorded by
 $\mathbb{P}^t$. Lines 17--21 determine the transmit node and select beam for
 $\mathbb{U}_t$. We first obtain the maximum achievable rate and corresponding beam
 for each user in $\mathbb{U}_s$ to serve $\mathbb{U}_t$, denoted by $R_t^i$ and
 $\varphi (b_t^i,l_t^i)$. Then, we select the user in $\mathbb{U}_{s}$ with the highest
 maximum achievable rate as the transmit node for $\mathbb{U}_t$, and the corresponding
 beam is recorded, which is the appropriate beam we referred to before. The algorithm is completed in line 22 by returning $\mathbb{U}_{\cal M}$,
 $\mathbb{P}^t$, and the selected transmit node and beam for each subset.

 The outer loop of lines 3--21 has $|\mathbb{U}|$ iterations. Each of the three inner
 loops, lines 5--7, lines 9--15 and lines 17--19, has at most $|\mathbb{U}|$ iterations.
 Moreover, the operations inside each inner loop impose at most the complexity on the
 order of $|\mathbb{U}|$. Therefore, the worst-case computational complexity of
 Algorithm~\ref{alg: multicast-D2D} is on the order of $\mathsf{O}(|\mathbb{U}|^3)$.

\subsection{Multicast Scheduling Algorithm}\label{S5-2}

 The proposed multicast scheduling algorithm iteratively allocates the multicast
 transmission for each subset into each phase until all the subsets are scheduled. We
 will denote the scheduled multicast transmission in the $k$th phase by $\mathbb{E}^k$.
 The pseudo-code of this multicast scheduling is given in Algorithm~\ref{alg:MSA}.
 Note that to meet the requirement of constraint (\ref{cons7}), only the user with the
 multicast data is able to serve other users. Thus, if user $i$ in subset $\mathbb{U}_s$
 is the transmit node for subset $\mathbb{U}_t$, the multicast transmission to
 $\mathbb{U}_s$ must be scheduled before the transmission to $\mathbb{U}_t$. In
 Algorithm~\ref{alg: multicast-D2D}, we obtain $\mathbb{U}_t$ after $\mathbb{U}_s$, and
 this order naturally meets constraint (\ref{cons7}). Therefore, we can simply schedule
 the transmissions to the subsets one by one by following the same order as recorded by
 Algorithm~\ref{alg: multicast-D2D}, as indicated in lines 4--6 of Algorithm~\ref{alg:MSA},
 which select the subset for the $k$th phase. The transmit node and beam determined by
 Algorithm~\ref{alg: multicast-D2D} for this subset are then used for the multicast
 transmission to this subset in the $k$th phase, as indicated in lines 7--8. As shown in line 7,
 there is only one multicast transmission for each phase, which is required by constraint (\ref{cons4}).
 The scheduling results for all the phases are outputted in line 9.

\begin{algorithm}[htbp]
\caption{Multicast Scheduling.}
\label{alg:MSA} 
 \textbf{Input}: $\mathbb{U}_{\cal M}$; $\mathbb{P}^t$, $s_t$ and $\varphi (b_t,l_t)$
  for each $\mathbb{U}_t \in \mathbb{U}_{\cal M}$; \\
 \textbf{Initialization}: $k$=0; \\
 \While {$k <|\mathbb{U}_{\cal M}|-1$}
  {
   $k$=$k$+1; \\
   Set $\mathbb{E}^k=\emptyset$; \\
   Find the $k$th transmission, $\mathbb{P}^k$; \\
   $\mathbb{E}^k=\{s_k \to \mathbb{U}_k\}$; \\
   Set the transmit beam for $s_k$ to $\varphi (b_k,l_k)$; \\
  }
 \textbf{Return} $\mathbb{E}^k$ for each phase.
\end{algorithm}

 The computational complexity of Algorithm~\ref{alg:MSA} is obviously on the order of
 $\mathsf{O}\big(|\mathbb{U}_{\cal M}|\big)$, where $|\mathbb{U}_{\cal M}|\le |\mathbb{U}|$,
 which is negligible compared with the computational complexity of
 Algorithm~\ref{alg: multicast-D2D}. Therefore, the computational complexity of our proposed
 multicast scheduling scheme MD2D, which consists of Algorithm~\ref{alg: multicast-D2D} and
 Algorithm~\ref{alg:MSA}, is on the order of $\mathsf{O}(|\mathbb{U}|^3)$.

\section{Performance Evaluation}\label{S6}

 This section evaluates our MD2D scheme. We also investigate the impact of the two thresholds in
 Algorithm~\ref{alg: multicast-D2D}, $r_{th}$ and $\theta_{th}$, on the achievable system
 throughput, energy consumption, and energy efficiency.

\subsection{Simulation Setup}\label{S6-1}

 In an mmWave small cell with $|\mathbb{U}|$ users, the AP is located in the
 center of a $20\,\text{m}\times 20\,\text{m}$ square area and the users are uniformly and
 randomly distributed in the area. After the bootstrapping program, we assume the network topology and location information of nodes have been collected by the AP, and the information will be updated periodically. During each frame, due to Gbps transmission rate in the mmWave band, the signalling overhead involving D2D path planning and transmission scheduling is small, and does not have a significant impact on system performance \cite{mao}. Besides, the difference in overhead for different schemes is small, and thus we mainly consider the transmission part for performance evaluation.

 We adopt the directional antenna model from IEEE 802.15.3c
 with a main lobe of Gaussian form in linear scale and constant level of side lobes \cite{chen_2}.
 The gain of the directional antenna in dB, denoted by $G(\theta )$, can be expressed as
\begin{equation}\label{eq15}
 G(\theta ) \! = \! \left\{ \!\! \begin{array}{cl}
  G_0 - 3.01 \cdot \Big(\frac{2\theta}{\theta_{-3{\rm dB}}}\Big)^2 , & 0^\circ \le \theta \le \theta_{ml}/2 , \\
  G_{sl} , & \theta_{ml}/2 \le \theta  \le 180^\circ ,
 \end{array} \right. \!\!
\end{equation}
 where the angle $\theta$ takes the value in $\big[0^\circ, ~ 180^\circ\big]$,
 $\theta_{-3{\rm dB}}$ is the angle of the half-power beamwidth. In the simulation, we adopt the four-level codebook, where the half-power
 beamwidth $\theta_{-3{\rm dB}}$ is equal to $15^\circ$, $30^\circ$, $45^\circ$ and $60^\circ$,
 respectively. The parameters of the simulated mmWave small cell are summarized in
 Table~\ref{tab:para-MD2D} \cite{tvt_own,JSAC_own}. For each experiment, we perform one hundred independent simulations
 and take the average of the results.

\begin{table}[tp!]
\begin{center}
\caption{Simulation Parameters of the simulated mmWave small cell.}
\label{tab:para-MD2D} 
\begin{tabular}{ccc}
\hline
\textbf{Parameter} & \textbf{Symbol} & \textbf{Value} \\ \hline
 System bandwidth & W & 2160 MHz \\
 Background noise &$N_0$& -134dBm/MHz \\
 Path loss exponent & $\tau$ & 2 \\
 Number of users & $|\mathbb{U}|$ & $5\sim 30$ \\
 Maximum Transmission power & $P_t$ & $30\sim 40$ dBm \\
 Time slot duration & $\Delta$ & 18 $\mu$s \\
 Efficiency of the transceiver design & $\eta$ & 0.5 \\
 Multicast data size & $D$ & $1\sim 10$ Gb \\\hline
\end{tabular}
\end{center}
\vspace*{-5mm}
\end{table}

 We adopt the
 following three performance metrics.

 1)~\textbf{Network Throughput}: The achieved multicast throughput of all the users in the
 network, expressed as
\begin{equation}\label{eq16}
 \text{NT} = \frac{|\mathbb{U}| \cdot D}{\sum\limits_{k=1}^K \delta^k\cdot \Delta}  ~~ \text{[b/s]} ,
\end{equation}

 2)~\textbf{Energy Consumption}: The total energy consumption of all the multicast
 transmissions, given by
\begin{equation}\label{eq17}
 \text{EC} = \sum\limits_{k=1}^K \frac{D}{R_k} \cdot P_t ~~ \text{[J]} ,
\end{equation}
 where $R_k$ is the transmission rate in the $k$th phase.

 3)~\textbf{Energy Efficiency}: The ratio of the achieved network throughput over the consumed energy,
 which is expressed as
\begin{equation}\label{eq18}
 \text{EE} = \frac{\text{NT}}{\text{EC}} = \frac{|\mathbb{U}|\cdot D}{\sum\limits_{k=1}^K \delta^k\cdot \Delta}
  \cdot \frac{1}{\sum\limits_{k=1}^K \frac{D}{R_k} \cdot P_t} ~~ \text{[b/s/J]} .
\end{equation}

 As reviewed in Section~\ref{S2}, there exists no previous work in the existing literature that
 joint exploits  D2D communications and multi-level codebook for improving multicast efficiency
 in mmWave based networks. In order to demonstrate the advantages of utilizing both D2D
 communications and multi-level codebook in our MD2D scheme, we compare our scheme with the
 following three multicast schemes.

 1)~\textbf{FDMAC}: In the FDMAC scheme, the AP sequentially transmits the multicast
 data to the users one by one using the finest-level beam. This is the baseline scheme which
 exploits neither multi-level codebook nor D2D communications \cite{mao}.

 2)~\textbf{MC}: In the multi-level codebook scheme, the multicast group is divided into
 different subsets, and the AP selects an optimal beam from the multi-level codebook to serve
 each subset. In this scheme, the multi-level codebook is exploited but D2D communications are
 not enabled. Through comparison with the MC scheme, the advantages of using D2D communications in our scheme can be observed.

 3)~\textbf{D2D}: In the D2D multicast scheme, D2D communications are utilized to improve the
 system performance, similar to the MD2D. However, for this D2D-only scheme, the finest-level
 beam is always used for each transmission, where only one user is served. Therefore, unlike
 our MD2D, this scheme does not utilize the multi-level codebook. Through the comparison with the D2D scheme,
 the advantages of our scheme due to the multi-level codebook are demonstrated.

 To evaluate our scheme under NLOS transmissions, we adopt the NLOS parameters in ~\cite{NLOS}, where the path loss exponent is equal to 3.01, and the shadowing effect is also considered. In the following performance evaluation, the results under NLOS transmissions are also presented.

\subsection{Impact of the User Partition Thresholds} \label{S6-2}

\begin{figure}[htbp]
\vspace*{-2mm}
\begin{minipage}[t]{1\linewidth}
\centering
\includegraphics[width=0.95\columnwidth]{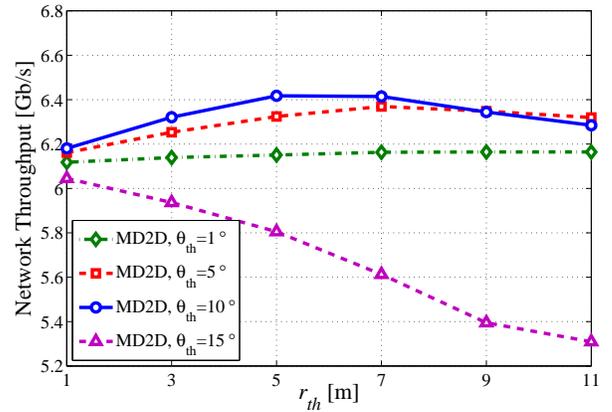}
\end{minipage}%
\vspace*{-3mm}
\caption{The network throughput achieved by our proposed MD2D given different radius and
 angle thresholds, given $\mathbb{U}=9$, $P_t=30$\,dBm and $D=1$\,Gb.}
\label{fig:threshold} 
\vspace*{-1mm}
\end{figure}

 Intuitively, the choice of the radius threshold $r_{th}$ and angle threshold $\theta_{th}$ in
 Algorithm~\ref{alg: multicast-D2D} of user partition and multicast path planning will seriously
 impact the performance of our MD2D scheme. For the system specified in Section~\ref{S6-1},
 Fig.~\ref{fig:threshold} plots the network throughput performance achieved by our MD2D with
 various radius and angle threshold values.

 In terms of the impact of angle threshold, too small or too large $\theta_{th}$ will degrade
 the achieved NT performance metric. More specifically, with a small angle threshold of
 $\theta_{th}=1^\circ$, only a small number of users located nearby can be allocated into a
 same subset, and MD2D will choose narrow beams to serve each subset. Consequently, the
 potential of the multi-level codebook is not fully exploited, and the number of transmissions
 increases, which degrades the network throughput. On the other hand, with a large angle
 threshold of $\theta_{th}=15^\circ$, many more users will be allocated into a same subset.
 MD2D will serve such a subset via a wide beam, and the transmission rate will decrease.
 Besides, some users served in this way would be much better served via more efficient D2D
 communications. For this system, it can be observed that with $\theta_{th}=10^\circ$, MD2D
 achieves the best performance.

 In terms of the impact of radius threshold, given a too small angle threshold of
 $\theta_{th}=1^\circ$, the influence of $r_{th}$ to the achieved NT performance metric is
 very small. By contrast, given a too large angle threshold of $\theta_{th}=15^\circ$,
 increasing $r_{th}$ degrades the achieved NT performance metric seriously. With the `optimal'
  angle threshold $\theta_{th}=10^\circ$, MD2D with $r_{th}=5$\,m achieves the best performance
 and, moreover, for $r_{th}$  between 5\,m and 7\,m, the network throughputs are all very good.
 Observe that in the case of $\theta_{th}=10^\circ$, when $r_{th}$ is small, the network
 throughput increases with the radius threshold. This is because more users are allocated
 into a same subset, and wide beams are able to serve these subsets simultaneously, which
 unleashes the potential of the multi-level codebook. However, when $r_{th}$ is large,
 increasing $r_{th}$ degrades the performance. This is because similar to the case of too
 large angle threshold, too many users will be allocated into a same subset which reduces
 the transmission rate and, moreover, D2D communications could not be exploited fully to
 improve the network performance. With the angle threshold $\theta_{th}=5^\circ$, the best
 choice of radius threshold is $r_{th}=7$\,m.

\begin{figure}[htbp]
\vspace*{-2mm}
\begin{minipage}[t]{1\linewidth}
\centering
\includegraphics[width=0.95\columnwidth]{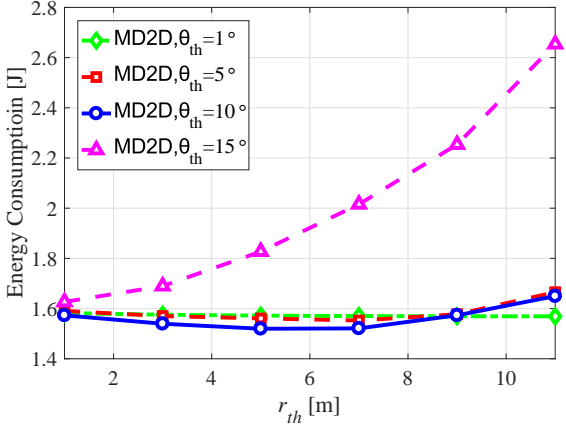}
\end{minipage}%
\vspace*{-3mm}
\caption{The energy consumptions achieved by our proposed MD2D given different radius and
 angle thresholds, given $\mathbb{U}=9$, $P_t=30$\,dBm and $D=1$\,Gb.}
\label{fig:ec_threshold} 
\vspace*{-1mm}
\end{figure}

 Besides, in Fig.~\ref{fig:ec_threshold}, we plot the energy consumptions achieved by our MD2D with different
 radius thresholds and angle thresholds. We can obverse that MD2D with angle threshold $\theta_{th}=10^\circ$ and radius threshold $r_{th}$  between 5\,m and 7\,m consumes least energy. For this system, higher throughput means that it can finish task scheduling faster, which causes that the system consumes less energy. According to the above analysis and simulation results, with the `optimal' angle threshold $\theta_{th}=10^\circ$, MD2D with radius threshold $r_{th}$  between 5\,m and 7\,m, achieves the best network throughput performance. Hence, MD2D with the same threshold ought to consume least energy, which agrees with the simulation results.



 For this system, it can be seen that the optimal choice of angle threshold and radius
 threshold is $\theta_{th}$ around $10^\circ$ and $r_{th}$ in the range of 5\,m to 7\,m.
 From this experiment, we may conclude that the angle threshold and radius threshold should
 be optimized according to the network environment, in order to maximize the achieved
 network throughput performance. Although the optimal threshold may be different under each case, the threshold can be selected to be optimized for the most of the cases. On the other hand, when the performance degrades too much in some cases, the system is assumed to adapt to the changes, and adjust the thresholds in our scheme.

\subsection{Comparison with Other Schemes} \label{S6-3}

 We now compare the NT, EC and EE performance of our MD2D scheme with those of the other three
 schemes, i.e., \textbf{FDMAC}, \textbf{MC} and \textbf{D2D}. The user partition thresholds
 of our MD2D, $r_{th}$ and $\theta_{th}$, are set to 6\,m and $10^\circ$, respectively, according
 to the investigation of the previous section.

\begin{figure}[htbp]
\vspace*{-1mm}
\begin{minipage}[t]{1\linewidth}
\centering
\includegraphics[width=0.95\columnwidth]{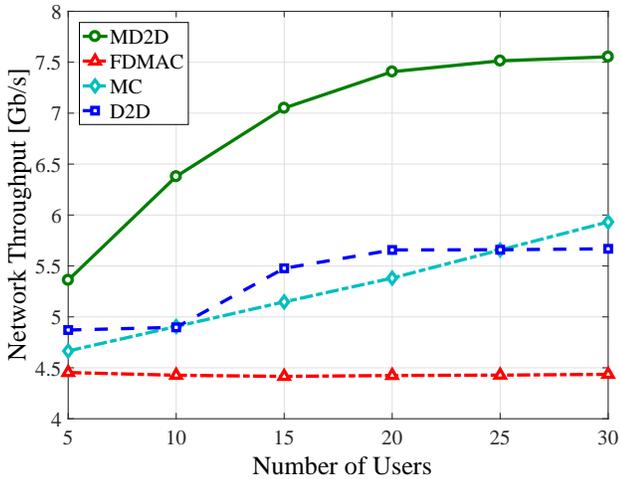}
\end{minipage}%
\vspace*{-3mm}
\caption{Comparison of the network throughputs as the functions of user number for four schemes under LOS transmission assumption, given $P_t=30$\,dBm and $D=1$\,Gb.}
\label{fig:throughput-compare} 
\vspace*{-1mm}
\end{figure}

\subsubsection{Network Throughput}

 Fig.~\ref{fig:throughput-compare} compares the NT performance metrics achieved by the
 four schemes for different numbers of users. As expected, the FDMAC scheme attains the worst
 performance and its NT metric remains constant with the increase of $|\mathbb{U}|$. By contrast,
 the NT metric of the MC scheme increases linearly with $|\mathbb{U}|$. With the increase
 of users, the MC scheme can exploit the multi-level codebook more effectively.
 As for the D2D scheme, its NT metric is relatively low when $|\mathbb{U}|$ is small. But when
 there are sufficient users in the network, the benefit of D2D communications becomes significant,
 leading to the considerable increase in the achieved NT performance. However, as $|\mathbb{U}|$
 increases further, its NT performance becomes saturated. The results of
 Fig.~\ref{fig:throughput-compare} also confirm that our MD2D scheme achieves the best performance
 among the four multicast schemes. For example, given $|\mathbb{U}|=5$, the MD2D scheme improves the
 network throughput by about 10\% over the second-best D2D scheme, while with $|\mathbb{U}|=30$,
 our MD2D scheme outperforms the second-best MC scheme by 27\%.

 \begin{figure}[htbp]
\vspace*{-1mm}
\begin{minipage}[t]{1\linewidth}
\centering
\includegraphics[width=0.95\columnwidth]{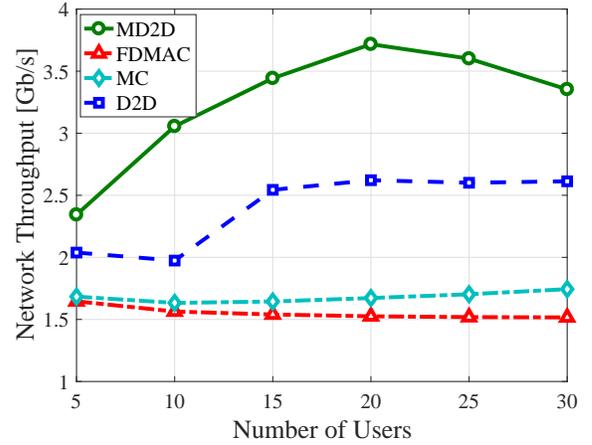}
\end{minipage}%
\vspace*{-3mm}
\caption{Comparison of the network throughputs as the functions of user number for four schemes under NLOS transmission assumption, given $P_t=30$\,dBm and $D=1$\,Gb.}
\label{fig:throughput-compare-nlos} 
\vspace*{-1mm}
\end{figure}

 In Fig.~\ref{fig:throughput-compare-nlos}, the network throughput comparison of our scheme and three other schemes with different number of users under NLOS transmission assumption is presented. The results show that our scheme performs best under NLOS transmission assumption. Compared with results under LOS assumption, MD2D scheme achieves lower throughput since links suffer higher propagation loss under NLOS transmission assumption.

\begin{figure}[htbp]
\vspace*{-1mm}
\begin{minipage}[t]{1\linewidth}
\centering
\includegraphics[width=0.95\columnwidth]{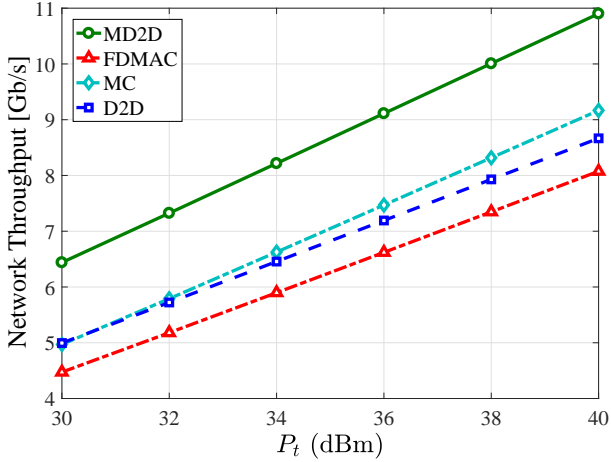}
\end{minipage}%
\vspace*{-3mm}
\caption{Comparison of the network throughputs as the functions of transmission power for four
 schemes under LOS transmission assumption, given $\mathbb{U}=9$ and $D=1$\,Gb.}
\label{fig:trasmissioinpower} 
\vspace*{-1mm}
\end{figure}

 Fig.~\ref{fig:trasmissioinpower} depicts the NT metrics as the functions of $P_t$ achieved by
 the four schemes. Since the $P_t$ is in dBm in the figure, the relationship between NT and transmission power is still consistent with Shannon law. As expected, the FDMAC scheme attains the worst performance, while our MD2D achieves the best performance. Specifically, the performance gap between our MD2D scheme and the second-best MC scheme
 increases from 1.5\,Gb/s to 1.9\,Gb/s as $P_t$ increases from 30\,dBm to 40\,dBm.

\begin{figure}[htbp]
\vspace*{-1mm}
\begin{minipage}[t]{1\linewidth}
\centering
\includegraphics[width=0.95\columnwidth]{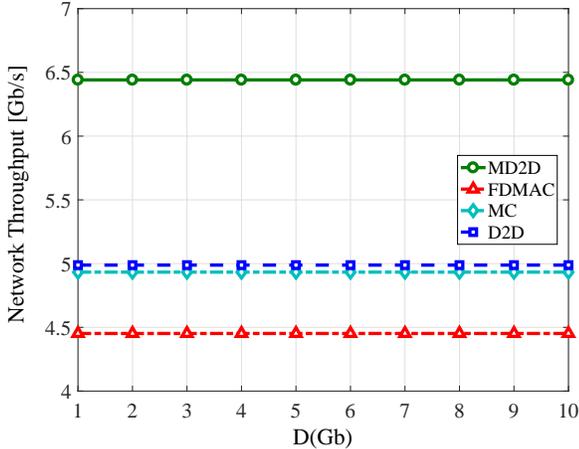}
\end{minipage}%
\vspace*{-3mm}
\caption{Comparison of the network throughputs as the functions of multicast data size for four
 schemes under LOS transmission assumption, given $\mathbb{U}=9$ and $P_t=30$\,dBm.}
\label{fig:throughput-D} 
\vspace*{-1mm}
\end{figure}

 Fig.~\ref{fig:throughput-D} compares the NT performance metrics achieved by the four schemes,
 given different multicast data sizes. Since the number of time slots required scales with the
 multicast data size, the network throughput remains constant when the multicast data size changes.
 Again the FDMAC scheme attains the worst performance and our MD2D achieves the best performance.
 The performance gap between our MD2D scheme and the second-best D2D scheme is 1.4\,Gb/s.

\begin{figure}[htbp]
\vspace*{-1mm}
\begin{minipage}[t]{1\linewidth}
\centering
\includegraphics[width=0.95\columnwidth]{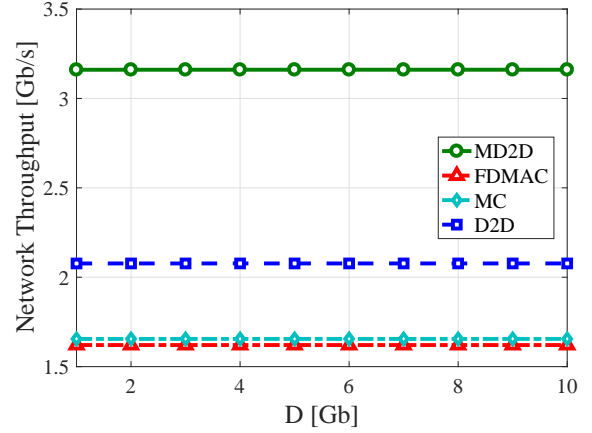}
\end{minipage}%
\vspace*{-3mm}
\caption{Comparison of the network throughputs as the functions of multicast data size for four
 schemes under NLOS transmission assumption, given $\mathbb{U}=9$ and $P_t=30$\,dBm.}
\label{fig:throughput-D-nlos} 
\vspace*{-1mm}
\end{figure}

 In Fig.~\ref{fig:throughput-D-nlos}, we plot the network throughput comparison with different multicast data sizes under NLOS transmission assumption. We can observe that our scheme achieves best, and still lower throughput is achieved due to higher propagation loss.

\subsubsection{Energy Consumption}

\begin{figure}[tbp]
\begin{minipage}[t]{1\linewidth}
\centering
\includegraphics[width=0.95\columnwidth]{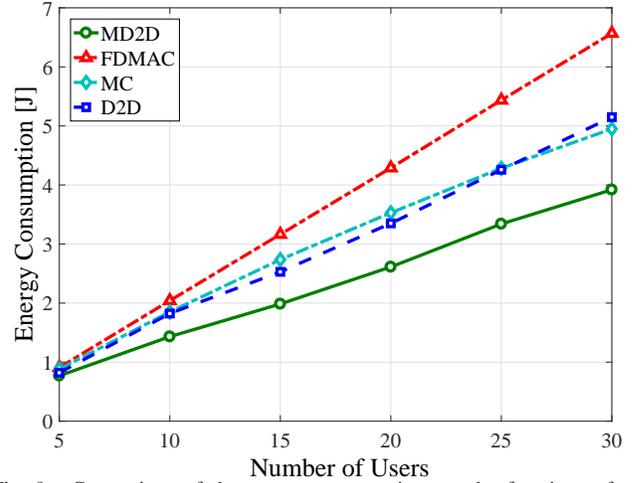}
\end{minipage}%
\vspace*{-3mm}
\caption{Comparison of the energy consumptions as the functions of user number for four
 schemes under LOS transmission assumption, given $P_t=30$\,dBm and $D=1$\,Gb.}
\label{fig:ec-compare} 
\vspace*{-4mm}
\end{figure}

 Fig.~\ref{fig:ec-compare} compares the energy consumptions of the four schemes under
 different numbers of users. Clearly, the energy consumption increases linearly with the number of
 users. Observe from both Figs.~\ref{fig:ec-compare} and \ref{fig:throughput-compare} that our MD2D
 scheme consumes the lowest energy consumption and achieves the highest network throughput. This is because higher throughput means that the system can finish the task scheduling faster, which causes that the system consumes less energy. For
 example, when the number of users is $|\mathbb{U}|=30$, our MD2D consumes 1\,J less energy, while
 increasing the network throughput by 1.6\,Gb/s, compared with the second-best MC scheme. This
 clearly demonstrates the significant benefits of jointly exploiting multi-level code book and D2D
 communications.

\begin{figure}[tbp]
\begin{minipage}[t]{1\linewidth}
\centering
\includegraphics[width=0.95\columnwidth]{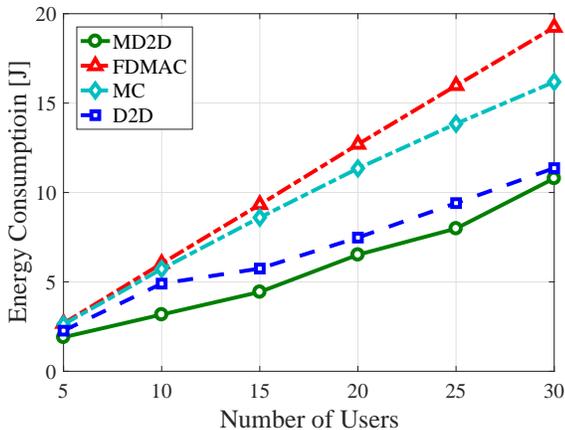}
\end{minipage}%
\vspace*{-3mm}
\caption{Comparison of the energy consumptions as the functions of user number for four
 schemes under NLOS transmission assumption, given $P_t=30$\,dBm and $D=1$\,Gb.}
\label{fig:ec-compare-nlos} 
\vspace*{-4mm}
\end{figure}

 Fig.~\ref{fig:ec-compare-nlos} plots the energy consumption comparison with different number of users under NLOS transmission assumption. From the results, we can observe that our scheme achieves lower energy consumption compared with other schemes. Compared with the LOS case, the energy consumption is much higher for all schemes since worse channel conditions under NLOS transmission assumption to complete the same multicast task.

\begin{figure}[htbp]
\vspace*{-1mm}
\begin{minipage}[t]{1\linewidth}
\centering
\includegraphics[width=0.95\columnwidth]{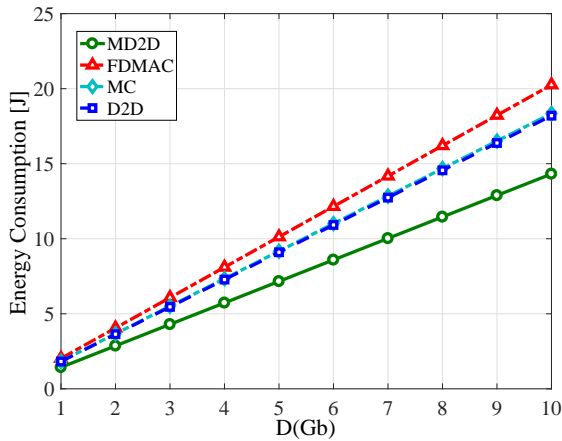}
\end{minipage}%
\vspace*{-3mm}
\caption{Comparison of the energy consumptions as the functions of multicast data size for
 four schemes, given $\mathbb{U}=9$ and $P_t=30$\,dBm.}
\label{fig:ec-compare-D} 
\vspace*{-1mm}
\end{figure}

 Fig.~\ref{fig:ec-compare-D} plots the energy consumptions of the four schemes given different
 multicast data sizes. The energy consumption increases with $D$ since the system needs more time to transmit data. Not
 surprisingly, our MD2D scheme consumes the least energy, while attaining the highest network
 throughput. Also observe that the gaps of energy consumption between our MD2D and other
 schemes increase with the multicast data size. In particular, given the multicast data size
 $D=10$\,Gb, our MD2D consumes 22\% less energy than the second-best D2D scheme.

\subsubsection{Energy Efficiency}

\begin{figure}[tbp]
\begin{minipage}[t]{1\linewidth}
\centering
\includegraphics[width=0.95\columnwidth]{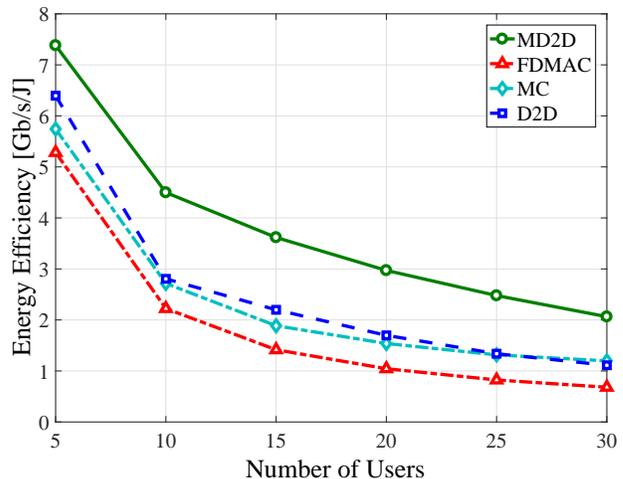}
\end{minipage}%
\vspace*{-3mm}
\caption{Comparison of the energy efficiencies as the functions of user number for four
 schemes under LOS transmission assumption, given $P_t=30$\,dBm and $D=1$\,Gb.}
\label{fig:ee-compare} 
\vspace*{-4mm}
\end{figure}

 The EE metric combines both the network throughput and energy consumption performance. In
 Fig.~\ref{fig:ee-compare}, we plot the EE metrics achieved by the four schemes given different
 numbers of users. With the increase of users, the traffic load of the network increases and
 the energy efficiency generally decreases, as can be seen from Fig.~\ref{fig:ee-compare}.
 Since network throughput affects energy consumption, and network throughput and energy consumption affect energy efficiency, as expected, MD2D achieves the highest energy efficiency performance because MD2D achieves the best NT performance and consumes the least energy among the four multicast schemes. In particular, given the number
 of users $|\mathbb{U}|=30$, our MD2D improves the energy efficiency by about 72\% compared
 with the second-best MC scheme.

\begin{figure}[tbp]
\begin{minipage}[t]{1\linewidth}
\centering
\includegraphics[width=0.95\columnwidth]{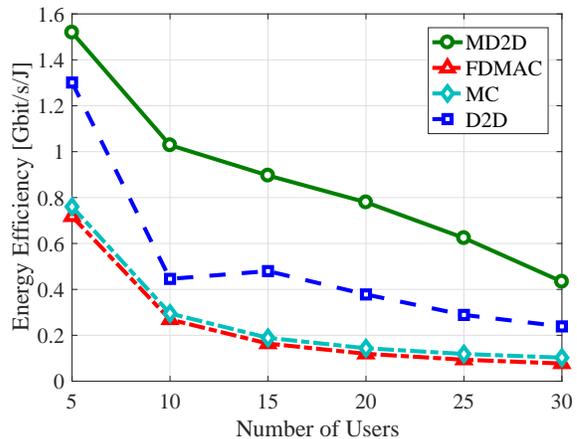}
\end{minipage}%
\vspace*{-3mm}
\caption{Comparison of the energy efficiencies as the functions of user number for four
 schemes under NLOS transmission assumption, given $P_t=30$\,dBm and $D=1$\,Gb.}
\label{fig:ee-compare-nlos} 
\vspace*{-4mm}
\end{figure}

 Fig.~\ref{fig:ee-compare-nlos} plots the energy efficiency comparison with different number of users under NLOS transmission assumption. We can observe that our scheme has the highest energy efficiency among all the schemes. Compared with the results under the LOS assumption, the achieved energy efficiency is lower since lower throughput and higher energy consumption under NLOS transmissions.

\begin{figure}[htbp]
\vspace*{-1mm}
\begin{minipage}[t]{1\linewidth}
\centering
\includegraphics[width=0.95\columnwidth]{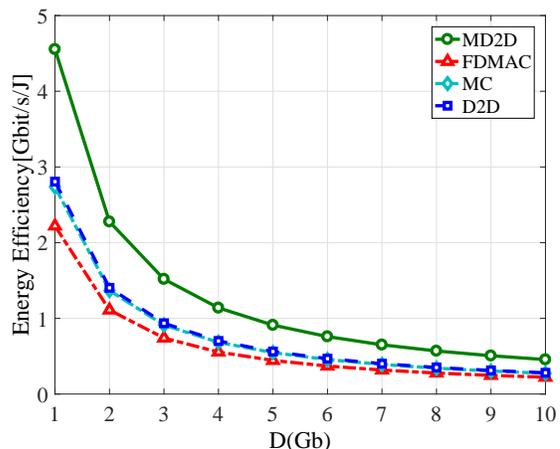}
\end{minipage}%
\vspace*{-3mm}
\caption{Comparison of the energy efficiencies  as the functions of multicast data size for
 four schemes, given $\mathbb{U}=9$ and $P_t=30$\,dBm.}
\label{fig:ee-compare-D} 
\vspace*{-4mm}
\end{figure}

 Fig.~\ref{fig:ee-compare-D} compares the EE metrics of the four schemes under different multicast
 data sizes. Since higher energy consumption is necessary for larger multicast data size, the energy
 efficiency generally decreases with the multicast data size. Not surprisingly, our MD2D achieves
 the highest energy efficiency among the four schemes. Compared to the second-best D2D scheme, our
 MD2D improves the energy efficiency by about 64\% and 66\%, respectively, given the  multicast
 data sizes  $D=1$\,Gb and $D=10$\,Gb.

\section{Conclusions}\label{S7}

 We have developed an efficient multicast scheduling scheme for mmWave small cells,
 which jointly exploits both D2D transmissions and multi-level antenna codebook to improve
 multicast efficiency. Our novel contribution has been twofold. Firstly, we have shown that the
 optimal multicast scheduling problem by jointly optimizing the utilizations of D2D transmissions
 and multi-level antenna codebook is NP-hard. Secondly, in order to obtain practical and efficient
 solution, we have developed a novel  multicast scheduling scheme, called MD2D. More specifically,
 in our MD2D solution, an efficient user-partition and multicast-path-planning algorithm partitions
 users in the multicast group into subsets and selects the transmit node for each subset. Then
 an effective multicast scheduling algorithm schedules the transmission for each subset into each
 transmission phase. Performance evaluation has verified that our MD2D multicast scheduling
 scheme significantly outperforms the other two multicast scheduling schemes relying on D2D
 communications and multi-level antenna codebook alone, respectively, in terms of network throughput
 and energy efficiency.
In the future work, we will analyze the relationship between the objective function and the parameters such as angle and radius thresholds in a theoretically way.


\bibliographystyle{IEEEtran}

\end{document}